# On the Cost of Security Compliance in Information Systems

David Hofbauer[1], Igor Ivkic[1,2], Silia Maksuti[3,4], Andreas Aldrian[5], Markus Tauber[3]

[1]Forschung Burgenland, Eisenstadt, Austria
[2]Lancaster University, Lancaster, UK
[3]University of Applied Sciences Burgenland, Eisenstadt, Austria
[4]Luleå University of Technology, Luleå, Sweden
[5]AVL LIST GmbH

**ABSTRACT**

The onward development of information and communication technology has led to a new industrial revolution called Industry 4.0. This revolution involves Cyber-Physical Production Systems (CPPS), which consist of intelligent Cyber-Physical Systems that may be able to adapt themselves autonomously in a production environment. At the moment, machines in industrial environments are often not connected to the internet, which thus needs a point-to-point connection to access the device if necessary. Through Industry 4.0, these devices should enable remote access for smart maintenance through a connection to the outside world. However, this connection opens the gate for possible cyber-attacks and thus raises the question about providing security for these environments. Therefore, this paper used an adapted approach based on SixSigma to solve this security problem by investigating security standards. Security requirements were gathered and mapped to controls from well known security standards, formed into a catalog. This catalog includes assessment information to check how secure a solution for a use case is and also includes a link to an estimation method for implementation cost. Thus this paper's outcome shows how to make Industry 4.0 use cases secure by fulfilling security standard controls and how to estimate the resulting implementation costs.

## 1. INTRODUCTION

Through the fourth industrial revolution, called Industry 4.0, which involves Cyber-Physical Production Systems (CPPS), boundaries of companies deteriorate. This is caused by the use of embedded systems and their exchange of data across the entire value chain [1]. Industrial Control Systems (ICS), for example machines with sensors for monitoring manufacturing and process automation within a production company serve as one of the basics for Industry 4.0. These machines have been traditionally built as stand-alone systems, not connected to the outside world and only accessible via point-to-point connection. Thus, maintaining these machines might cause an overhead, since it often includes to physically move them in order to be able to establish a connection. To improve the maintenance, process a transformation to Industry 4.0 [2] could enable remote access, which allows servicing the machine from outside.

However, by no longer isolating ICS, the gates for numerous cyber-attacks are opened and thus organizations must understand the potential risks for that. Before opening such a system to the outside world, it is important to know i) what to do in order to be secure and ii) how much it costs to do so. Thus, being able to provide a secure remote access is as important as providing a secure end-to-end communication in Industry 4.0. Since for the latter a lot of research work is done (e.g. in [3], [4], [5], [6]) this paper provides an approach how to make a use case secure for Industry 4.0 environments based on security standards and what costs for this security implementation can be estimated.

To achieve this, the approach adapted in this paper is based on SixSigma. Therefore, requirements are gathered and security standards are used to find appropriate controls addressing these requirements. Afterwards these outcomes are mapped into a requirements and controls catalog, which also includes assessment of controls and cost estimation for the implementation of controls. Through an experimental evaluation the usage of this approach's outcome is done to show it's applicability and implementation options. The method was adapted with other use cases in mind, which can be used for future work. The contributions of this paper are thus twofold:

- Firstly, an approach of how to design a secure Industry 4.0 application by fulfilling security standard controls and
- secondly, an approach for estimating the implementation costs to build such an application

Both should help building and implementing a cost aware and secure application for Industry 4.0. The remainder of the paper is structured as follows: Section 2 shows related work concerning the most important topics regarding this paper's method and how this work can be used or enhanced. Next, section 3 describes the methodology by explaining in detail each step of the used SixSigma approach. Based on that, section 4 shows in a step by step evaluation how the presented approach can be used to provide security and estimate the resulting costs for an industrial use case. Finally, in section 5 we summarize the approach including the evaluation results and give an outline for possible future work.

## 2. RELATED WORK

Brettel et al. [1] have investigated the topic Industry 4.0 in general. They have analyzed 8 different research journals concerning the topics individual production, end-to-end engineering in a virtual process chain and production networks through cluster analysis and also held face to face interviews with industry managers. Their results showed from a managerial viewpoint why companies adapt or refuse Industry 4.0 technologies. Their outputs should also help decision makers to assess if transformation to Industry 4.0 within their companies is needed. Further general work [7] has also investigated general Industry 4.0 aspects, more precisely design principles for Industry 4.0 scenarios. Through a quantitative and qualitative text and literature analysis and review, their outcome was the provision of design principles to create a common understanding of the term Industry 4.0 for reasonable scientific discussion and also through a case study the paper can help identify potential use cases for Industry 4.0. Both papers are used to explain basics of Industry 4.0 and also to help identifying important general topics. Our work will also include general explanation for Industry 4.0 and enhance it by prensting an actual Industry 4.0 use case.

Bicaku et al. [3] have proposed a solution to investigate automated standard compliance to assure that individual Industry 4.0 components are secured to interoperate. The basis of that are given sets of security and safety requirements from which measurable indicator points are derived. An initial approach to automate such assessment when components are inter-operating with each other is done by using a monitoring and standard compliance verification framework. Our work builds on this effort of checking security standard compliance and extends it by evaluating a use case scenario for Industry 4.0 by investigating corresponding requirements and thus deriving standards, controls and costs for implementing secure Industry 4.0 applications.

Schmittner et al. [8] have investigated the combination of safety and security for trustworthy cyber-physical systems and explain that in the past safety and security were handled as separate issues but now with Industry 4.0 and the interconnection of each device, they must be handled as one part. This combination needs new concepts, techniques and tools, which this special issue addresses. This paper considers this relation in mind and focuses on addressing such security issues.

The papers of the BMWi [5], [6] aim to define a simple and overall position about basic requirements, security challenges and diverse approaches for secure communication in Industry 4.0 environments. They also give a deeper look at the C-I-A triangle (Confidentiality-Integrity-Availability) in Industry 4.0 and authenticity. Our work makes use of these results, especially of CIA and authenticity for deriving requirements and thus standards, controls and costs.

In our previous work [9] we have introduced a new method of measuring the costs of cyber security in smart business. It presents an initial evaluation of the costs of cyber security using a high-level process flow based on Six Sigma. The limitations of this work were that a fictional use case was used, the cost analysis was performed only during design time and it used only a monetary cost metric (Euro). Therefore, this work extends these methodologies of measuring costs of security by evaluating and adapting the approach to the remote access requirements and controls evaluation taxonomy. Furthermore, the derivation of costs can also be considered and adapted to the costs in this work.

In [10], Yee provides a summary of related work regarding security metrics by establishing the argument that many security controls and metrics exist, but most of them are not effective or necessary. Afterwards, a definition of possible and not possible controls is provided and furthermore described what the difference between conventional and scientifically/reliably-based security controls is. The procedure of this work can serve for the identification of controls for this paper and extends it by finding usable possible "good" controls for the catalog.

Strobl et al. [11] have identified threats, vulnerabilities and their impact of connected cars in their paper. The aim of the paper was to identify blocks of established technologies in a connected car and to consolidate the corresponding threat and vulnerability catalogs relevant for the individual constituent components. These findings are used to estimate the impact on specific system components and subsystems to identify the most crucial components and threats including cloud computing. Our paper was also motivated by this approach of gathering controls for the collection of controls for the requirements and controls catalog including costs. Although the topics vary in our work, the build-up of the catalog can be used as a possible example for the requirements and controls catalog and extends it by adding new approaches.

Specifying security requirements is an important task to develop critical information systems. Several standards help to handle security requirements. Thus, Mellado et al. [12] present a common criteria centered and reuse-based process dealing with security requirements at an early stage of e.g. software development in a systematic and new way. This is done by providing security resource repository and by integrating them in the early stages of the software lifecycle, unifying concepts of requirements engineering and security engineering. So, they show a standard-based process that deals with the security requirements at early stages of software development but do not concern security and cost in Industry 4.0. In contrast to this work this paper shows an approach of standard-based security requirements engineering actually providing a catalog with requirements adapted to a use case with an intuitive way, so it could be reused and also integrating a cost estimation method.

In this work [13] a systematic review of security requirements engineering is performed, which is done by a systematic review of existing literature concerning security requirements engineering. This review was used to show that there are too less of these important reviews and to provide a framework/background, in which to appropriately position new research activities. Therefore, this reference provides a summary of all existing information about security requirements in a thorough and unbiased manner. Their main contribution includes precision and reliability of the information and the results obtained. This paper will examine such security requirements engineering sources and will provide a way based on SixSigma to come to security requirements and concerning costs through the help of standards.

## 3. RESEARCH METHODOLOGY

This paper focuses on fulfilling security standard controls as well as on the cost for implementing these. For this, the most important requirements and security standard controls for a use case are evaluated and based on that the needed implementation costs are estimated. An approach on how to derive these necessary requirements and controls was formed by asking the questions:

1. What does security in Industry 4.0 mean? – Based on our experience, it means to be compliant to security standards. Thus, requirements are necessary to define what needs to be secured.
2. How can requirements be fulfilled? – By using controls from security standards (= measures, which can be used to assess if something, in this case, is secure).

The outcome should be a requirements and controls catalog, which should present the information about security standard controls in such a way that the costs for implementing them can be derived. For this catalog and cost derivation, a taxonomy is needed to develop a method, which can be easily reused. A simple taxonomy can be derived from Six Sigma. The Six Sigma approach was developed by Motorola in the 1980s and is one of the foremost methodical practices for improving customer satisfaction and business processes. The idea behind the approach was to remove causes of errors when detected before they lead to defects in a product or service, which leads to more cost-efficiency. This is achieved by the Six Sigma DMAIC approach, which stands for Define, Measure, Analyze, Improve, Control. This approach can be used for many use cases, e.g. identifying cloud security risks [9] or in this case for gathering security controls. Thus, the steps for this approach are shown in Figure 1, in which the numbers in brackets () are used to describe exactly what every step means below the figure.

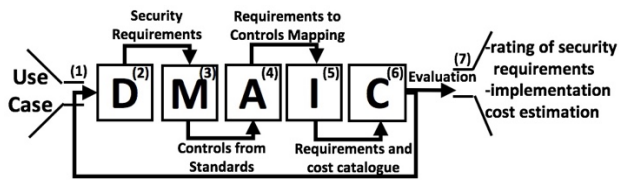

*Figure 1: Methodology based on SixSigma*

The goal of this methodology is to put an Industry 4.0 security use case through a hopper to start the DMAIC approach (1). Afterwards the application wanders through every step of the approach:

**D**efine (2): Identification of the requirements.

**M**easure (3): Based on the requirements, finding security controls from existing security standards.

**A**nalyze (4): Mapping of requirements to controls and including it in a catalog. A set of controls, addressing a specific requirement, is extracted from each standard, e.g. to address requirement 1, control 1 and 2 must be implemented. Furthermore, similar controls per requirement are summarized to control groups. For instance, for guaranteeing a secure connection, two different standards might suggest using a specific encryption algorithm as a security control. The catalog combines these similar controls from the standards in one group.

**I**mprove (5): To improve the catalog from the previous step, this phase provides guidance for how to assess implementing these security controls also including an equation for measuring the resulting costs.

**C**ontrol (6): Finally, the control phase uses the requirements and controls catalog to find possible solutions for the presented use case. The goal here is to show i) how well the solutions support the security controls from the used standards and ii) how much it costs to implement missing security controls.

After finishing these steps, the outcome is a rating of importance for security requirements as well as a cost implementation estimation and thus a secure and cost aware solution (7). If, during the Control phase no appropriate solutions were found, the DMAIC approach can be done again (1), which is explained in detail in the Control part of section 4. In this paper only one iteration is done because appropriate solutions were found during the Control phase. To show the described methodology in action, the next section presents a remote-access use case concerning Industry 4.0 and its iteration through the defined SixSigma approach.

## 4. EVALUATION

In the last section, we presented an approach for evaluating a use case through the help of defined taxonomy based on SixSigma to check if it fulfills security controls and also to estimate possible implementation costs. In this chapter, we will apply the defined methodology step by step, by first introducing the use case and then put it through each DMAIC step.

**Use Case**
In the following section, we demonstrate the first step of our approach through an industrial use case. The use case at the moment concerns measurement devices, which are connected to the company's local network with strict access rights. Currently, if a measurement device needs maintenance, it has to be serviced by accessing it via a point-to-point connection, which is a long lasting and unnecessarily complex process. Thus, the company decides to add to their measurement devices a remote access functionality. The new remote access use case has the following two basic goals: (i) providing an alternative way of accessing the device, and (ii) enabling quick and qualified support and maintenance from outside the company. Through the application of this remote access from outside, new security challenges arise, which this paper addresses through adding a check for standard controls implementation and its cost estimation.

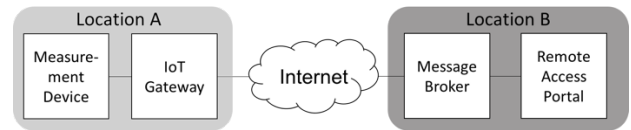

*Figure 2: Use Case setup*

Figure 2 shows the setup of the use case. On the left, there is Location A where the measurement device is in place. The right side of the picture represents another "outside" Location B, where the remote access to the device is needed. The whole solution is based on a message broker where location A is the publisher and location B is the subscriber. In Location A, the measurement device is only accessible through the customer's local area network therefore a remote access through an IoT Gateway is needed. The IoT Gateway can be seen as a programmed, intelligent switch, which can connect to the internet and establish an encrypted connection to a message broker in another location. A remote access server with a remote access web portal acts as a subscriber to the topic of the IoT Gateway, through which Location B can now directly access the device.

**Define**
The first step of the DMAIC approach is now used to identify requirements. Therefore, requirements were gathered based on CIA (Confidentiality, Integrity, Availability) [5] and extended with necessary requirements based on a use case and industry (which we call CIA extension) but also held on a general level with other use cases in mind. The six identified requirements were: Identification & Authentication [IA], Data Integrity [DI], Data Confidentiality [DC], Encryption [EC], Availability [AV] and Communication Channels [CC].

**Measure**
The next step was to identify controls from appropriate standards based on the requirements and the use case. These can of course vary from use case to use case or company. For the catalog, more than 5 Standards were found most appropriate. These standards have a high international reputation and therefore fit perfectly for the catalog. For the described Industry 4.0 remote access use case the following standards were used (split up in three parts caused by same controls specification): IEC 62443-3-3 [14], ISO/IEC 27000 series (27001 [15], 27002 [16], 27017 [17], 27018 [18]) and NIST SP (800-53 [19], 800-82 [20]). Based on these standards, controls related to the remote access use case were extracted.

**Analyze**
In this phase, requirements, controls and standards were mapped into a catalog and grouped in control groups that contain similar controls. Figure 3 shows the build up of the catalog: gathering requirements by the use of the in section 3 explained CIA extension and mapping controls of standards to them. The look of the catalog will be shown in Table 1 in the Improve phase.

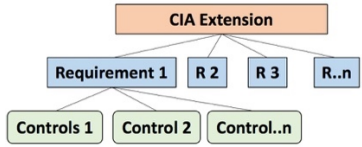
*Figure 3: Buildup of the requirements and controls catalog*

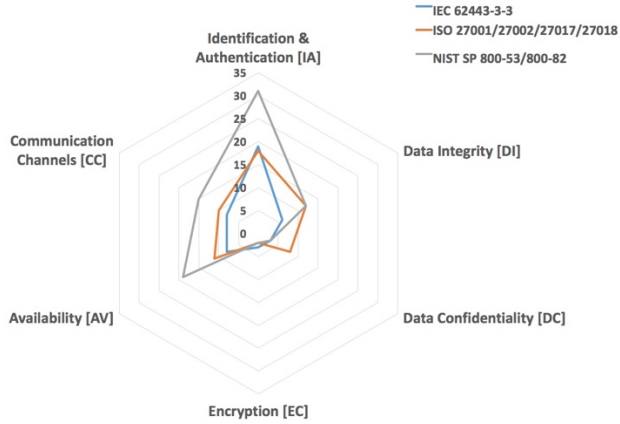
*Figure 4: Controls identified per requirement and standard*

Figure 4 shows the distribution of controls within the requirements and controls catalog across the standards. The figure shows the number of controls for each requirement and standard. Through this figure we can identify that e.g. the NIST standards have the highest number of controls concerning [IA] (thin grey line 0-35) whereas for [EC] all standards have a very low number of controls. Nevertheless, [IA] heavily depends on [EC] so dependabilities also need to be considered when implementing controls. But overall it is possible to say that identification & authentication have the most controls. Therefore, the most important requirement and controls that need to be considered for the whole remote access, also in combination with the underlying infrastructure and every part of it is [IA] and thus needs to be well considered before implementation. It also shows that NIST standards have the highest number of controls on most requirements. This Figure 4 is an important output of the paper because it shows a company what is most important for security, in this case, for the remote access use case.

**Improve**
The following Table 1 shows the build-up of the requirements and controls catalog, which also includes the mapping of requirements and controls from the Analyze phase. Moreover, it adds the assessment column and afterwards in equation (1) to (3) it shows the cost estimation method. The original catalog includes over 120 controls and spans over more than 20 pages, which can thus be not included in the paper. Nevertheless, Table 1 should give an overview of how the catalog looks like by showing the control groups for one requirement, Identification & Authentication. The controls for each requirement are grouped to control groups (column ID), which have the same/similar content (column Control IDs shows which controls are in this group). The assessment column then explains what these controls are about and how to assess if a used technology/application to solve a use case implements these security standards controls or not. The build-up of the table is as follows:
- Requirement (Req.)
- ID within the catalog (ID) = Control group ID
- concerning Controls IDs (Control IDs)
- Assessment information (Assessment)

| Req. | ID | Control IDs | Assessment |
|---|---|---|---|
| IA (Req. n) | 1 | [IEC-1] [IEC-2] [IEC-3] [IEC-4] [IEC-6] [IEC-8] [ISO-02-4] [ISO-02-5] [ISO-02-6] [ISO-02-8] [ISO-02-10] [NIST-53-1] [NIST-53-2] [NIST-53-4] [NIST-53-5] [NIST-53-18] [NIST-53-22] [NIST-53-23] [NIST-53-31] | Account & access management, including rights, duties, least privilege and access control. To address these controls following must be implemented in the use case to be in accordance with the standards: -Unified policies -Passwords, tokens, biometrics, etc. -General account management enabled (grouping, least privilege...) |
| IA | 2 | [17 controls] | |
| IA | 3 | [5 controls] | |

*Table 1: Extract from requirements and controls catalog*

Through the assessment column the requirements and controls catalog can thus serve as a controls fulfilment checklist as well as implementation recommendation guide. When assessing use case solutions with the catalog, a point system per control group should be used: Fully applicable gives 1 point, partly applicable 0,5 points and not applicable 0 points. The Control phase will show this assessment in detail.

Next, the challenge was to extend Table 1 with a column for an estimation of possible implementation cost. Therefore, the steps for the cost estimation are shown in equations (1), (2) and (3).

$$ct = \sum_{i=1} c_i \qquad c_i \in \{1\} \qquad (1)$$

$$ctMax = max(\{ct_1, ct_2, \ldots, ct_n\}) \qquad (2)$$

$$IE = \frac{ct}{ctMax} \qquad (3)$$

- $ct$    count = number of controls per control group
- $c_i$    control
- $ctMax$    count maximum = highest number of controls over all control groups of one requirement
- $IE$    Implementation Effort, result of equation

The following example Table 2 shows the steps how to come to the estimation of the implementation effort with the example of control group 1, 2 and 3 from requirement [IA] (see Table 1):

| | *Control group 1* | *Control group 2* | *Control group 3* |
|---|---|---|---|
| 1 | $ct_1 = \sum_{i=1} 1 = 19$ | $ct_2 = \sum_{i=1} 1 = 17$ | $ct_2 = \sum_{i=1} 1 = 5$ |
| 2 | $ctMax = max(\{19, 17, 5\}) = 19$ | | |
| 3 | $IE_1 = \frac{19}{19} = 1$ | $IE_2 = \frac{17}{19} = 0,89$ | $IE_3 = \frac{5}{19} = 0,26$ |

*Table 2: Cost estimation evaluation example*

| Req. | Control group ID | Nr. of controls per group (ct) | Implementation Effort (IE) |
|---|---|---|---|
| [IA] | 1 | 19 | 1 |
| | 2 | 17 | 0,89 |
| | 3 | 5 | 0,26 |

*Table 3: Cost estimation results*

To illustrate the cost derivation method results, Table 3 shows all control groups for one requirement, which is in this case again Identification & Authentication. The build-up of this table is thus almost the same as Table 1 but instead of every Control ID it is the number of Control IDs (*ct*) and instead of the assessment it contains the implementation effort (*IE*). Equation (1), (2) and (3) above show how the implementation effort is derived. Table 2 shows how the equations are used: Every number of controls per control group (*ct*) is divided through the highest number of controls over all three groups (*ctMax*), which is in this case 19 from control group 1, which leads to the implementation effort column of Table 3. Control group 3 in this case has the lowest implementation costs with 0,26, whereas control group 1 and 2 have high implementation costs with its implementation effort 1 and 0,89. The definition of high or low has to be done by the person or company itself that examines a certain use case because implementation effort can vary a lot. This method was used because implementing more controls means more e.g. time or working hours, which lead to higher costs. The method does not bring an exact cost estimation for implementation costs but it should give an overview and a feeling of how high or low the implementation of certain controls will be. This estimation of costs can be done with every requirement from the catalog. Figure 4 shows on an overall scale that Identification & Authentication has the most controls, which has through this approach the highest implementation costs because like Table 3 shows: *more controls = higher cost* based on the assumption that the costs for implementing the controls are the same.

**Control**
In this section, the Control phase of the SixSigma DMAIC approach is shown. It was done by evaluating possible platforms for the implementation of the remote access use case by the help of the requirements and controls catalog to show its applicability and to get the best security standard controls implemented solution. Based on the remote access use case and gathered requirements, 5 possible platforms as solution for it were found by using a research taxonomy: The platforms (i) should have at least one security certification, (ii) should already be used in Industry 4.0 use cases (two or more references), (iii) must already include on the website information regarding authentication, encryption & user management, which are key parts of the requirements and (iv) search keywords must include at least 2 of these keywords: remote access, IoT, Industry 4.0.

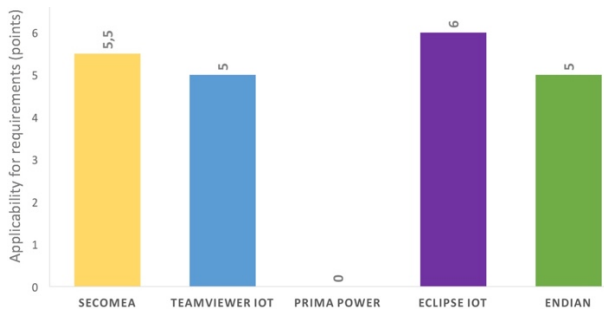

*Figure 5: Overall platform evaluation results*

At first these platforms were assessed against the catalog on a superficial, non-implementation level just with the documentation for these solutions. Figure 5 shows the applicability of the solutions for the use case per requirement (done with the point assessment from explained in the Improve phase). Next to the other found solutions (Secomea [21], Teamviewer [22] IoT, Prima Power Remote Care [23] and Endian [24]), Eclipse IoT [25] was the only fully applicable solution, which met the requirements and was then experimentally evaluated in detail.

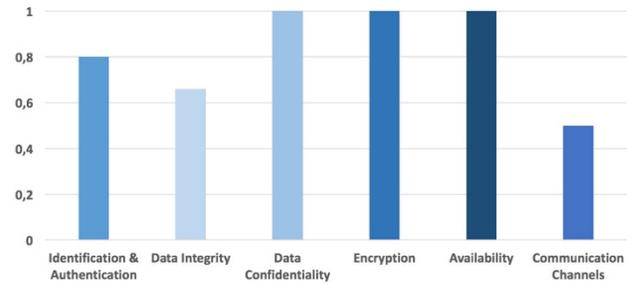

*Figure 6: Requirements applicability on Eclipse IoT from 0-1*

Eclipse IoT's solution (Kapua and Kura applications for remote access) were then implemented and again in detail evaluated against the requirements and controls catalog. The main outcome should then show that the overall evaluation must have rather the same outcome than the detailed one concerning one specific platform. The point evaluation for Figure 6 is the same as the one used for Figure 5. But here the control groups itself have been investigated in more detail through a setup, therefore the data ranges more. The most important security requirement of the platform is according to Figure 4 Identification & Authentication, which Eclipse IoT nearly fulfills except the policies, which need to be established by the company itself in advance, thus not part of this evaluation. Also, Encryption is well-included in Eclipse IoT, which is a very important requirement because it is also a part of all other gathered requirements. All 6 requirements are partly or fully fulfilled and have about 5 out of 6 achievable points, which is nearly the same outcome of the overall evaluation and thus shows the applicability of the requirements and controls catalog as a checklist. However, if the goal is to fulfill all requirements, the outcome of this phase provides the following three implementation possibilities: (i) if the missing parts have to be developed, the resulting costs could be estimated by using the presented method from the Improve phase; (ii) two or more solutions, which by themselves only partly fulfill the requirements, might fulfill all requirements when being combined; (iii) in case (i) and (ii) do not apply, another iteration of the DMAIC process (including other standards) could provide additional solutions.

**Results**
The main contribution of the paper is the approach shown in Figure 1. Based on this general approach, Industry 4.0 use cases can be assessed, security controls found and resulting implementation cost evaluated. In this regard the approach provides the following two main outcomes: (i) Figure 4 could be used to rate the requirements by their importance, where more controls are considered more important than less. (ii) This rating further leads to equations (1), (2) and (3), which allow to estimate the costs for implementing the identified security controls.
Further results include Table 1, which shows the catalog for checking if an application fulfills certain controls. Also, the Control phase shows results by presenting the applicability of the catalog on applications and also by showing appropriate solutions for the remote access use case implementation.

## 5. CONCLUSION & FUTURE WORK

This paper focuses on an approach on how to build secure and cost aware Industry 4.0 application. The first focus for achieving this was on investigating security. Based on the SixSigma DMAIC approach, security requirements were defined and controls in standards, which show how to achieve these, were gathered. Then a requirements and controls catalog regarding an Industry 4.0 remote access use case was formed. This catalog should help implement secure Industry 4.0 applications through the check if solutions fulfill controls and how much controls need to be implemented to be secure. Results included the rating of the most important requirements.

The second focus of the paper was on costs for implementing the most needed controls. Therefore, an approach of deriving implementation costs based on controls identified per control group within the catalog was presented. This result gives an implementation cost estimation to get an overview of what it will appropriately cost when implementing these. To show the applicability of the catalog and thus the Control phase of DMAIC, an evaluation was done with remote access applications to check on an overall and detailed level the applicability of this derived catalog. The outcome was a platform for the use case, which is the most secure one according to the catalog check.

Future work in general includes the adaptation of the catalog to different scenarios by expanding it or adding more standards. Further future work would be an automatic assessment of solutions with the requirements and controls catalog to expand e.g. existing work, which investigates automated standard compliance for monitoring [3]. It would also be interesting to do a more detailed implementation of the solutions with appropriate underlying infrastructure in a real Industry 4.0 environment.

## 6. AKNOWLEDGMENT

Research leading to these results has received funding from the EU ECSEL Joint Undertaking under grant agreement n737459 (Productive4.0) and from the partners national programs/funding authorities and the project MIT 4.0 (FE02), funded by IWB-EFRE 2014 - 2020 coordinated by Forschung Burgenland GmbH